\documentclass[a4paper]{PoS}
\usepackage{cite}
\usepackage{amssymb}
\usepackage[T1]{fontenc}
\usepackage{times}
\usepackage{mathptmx}
\usepackage{graphicx}
\usepackage{url}
\usepackage{esvect}
\newcommand{\pt}{p_{\mathrm{T}}}
\newcommand{\ptgen}{p_{\mathrm{T,jet}}^{\mathrm{gen}}}
\newcommand{\ptjet}{p_{\mathrm{T,jet}}}

\newcommand{\snn}{\sqrt{s_{\mathrm{NN}}}}
\title{Performance of the ALICE secondary vertex b-tagging algorithm in p-Pb collisions}
\ShortTitle{ALICE secondary vertex b-tagging algorithm}

\author{\speaker{Luk\'a\v{s} Kram\'arik} on behalf of the ALICE collaboration\\
        Department of Physics\\
        Faculty of Nuclear Sciences and Physical Engineering \\
        Czech Technical University in Prague\\
        B\v rehov\'a 7, 115 19 Prague 1, Czech Republic\\
        E-mail: \email{lukas.kramarik@fjfi.cvut.cz}}

\abstract{%
The hot and dense nuclear matter, that is produced in heavy-ion collisions, could be studied by jets originating from beauty quarks. In-medium energy loss of these quarks provides information on several properties of the quark-gluon plasma, produced in ultra-relativistic heavy-ion collisions. Reconstructed jets are powerful tools, since they offer access to kinematics of these hard-scattered partons.
Beauty hadrons are specific for their long lifetime, large mass and large-multiplicity decays.
Due to the long lifetime beauty hadrons decay at displaced secondary vertices.
In the ALICE experiment, secondary vertex properties are used to tag b-jets. The study of Monte Carlo based performance of the b-tagging algorithm for charged jets in p-Pb collisions at \mbox{$\snn=5.02$ TeV} is discussed in proceedings.}

\FullConference{54th International Winter Meeting on Nuclear Physics\\
		25-29 January 2016 \\
		Bormio, Italy}
\begin{document}

\section{Introduction}
\label{intro}
Heavy quarks (charm and beauty) are mainly produced in hard scatterings in the initial stage of hadronic collisions. They travel through the hot and dense QCD matter, Quark-Gluon Plasma (QGP), created in ultra-relativistic heavy-ion collisions, and lose energy via gluon radiation and elastic collisions.
The energy loss in the QGP depends on the colour charge and mass of the traveling parton. In case of the gluon radiation, the energy loss of light quarks and gluons is expected to be larger than for heavy quarks  \cite{Dokshitzer,Armesto}, so the transverse momentum spectra of heavy-flavour hadrons are expected to be less modified with respect to pp collisions. For low transverse momentum, collisional energy loss may also be important for heavy quarks in the QGP.
Reconstructed jets, coming from heavy-quark fragmentation, are a powerful tool to study the properties of the QGP. For quantitative studies of the QGP properties (e.g. density and transport coefficient) in Pb-Pb collisions, cold nuclear matter (CNM) effects, accessed with p-Pb collisions, are required.

The ALICE (A Large Ion Collider Experiment) detector \cite{ALICE} has great capability to reconstruct charged particles in a wide transverse momentum range, that are further used to reconstruct jets. Our goal is to tag jets coming from the fragmentation of heavy b quarks. Different tagging algorithms, based on properties of beauty-hadron decays, are studied in ALICE.
At the LHC (Large Hadron Collider) such studies are done in CMS \cite{CMS_frac,CMS_frac_pPb,CMS_Pb} and ATLAS \cite{atlas}. In CMS, b-jets in p-Pb collisions are studied for transverse momentum larger than 50~GeV/$c$. In ALICE, smaller transverse momentum of the jets is accessible. 
In this proceedings, the Monte Carlo (MC) simulation-based study  of the performance of the b-jet tagging algorithm exploiting displaced secondary vertex topologies for p-Pb collisions at $\snn=5.02$ TeV is presented. 

\section{Analysis details}
\label{sec1}
In this study p-Pb collisions at ${\snn} = 5.02 $ TeV are considered and simulated with combination of event generators PYTHIA6 (Perugia 2011 tune) \cite{PYTHIA} for pp collisions, and HIJING \cite{HIJING} for the underlying Pb background.
Collisions of protons at $\sqrt{s} = 5.02 $ TeV, simulated with a Lorentz boost (to get the same rapidity shift as in p-Pb collisions), are used to study the detector response for jets without underlying event (anything but the hard scattering). The response of the ALICE detector is simulated with GEANT3 \cite{GEANT}.

For the b-jet tagging the analysis steps are as follows:
 a selection of the tracks suitable for b-tagging, jet reconstruction using these tracks and underlying background subtraction under the jet area, tagging of b-jets with a particular algorithm and, finally, corrections of the jet spectrum. Two main corrections of the raw spectrum of b-tagged jets 
have to be made. One of them is unfolding (for detector effects) and the other is the correction for tagging efficiency and purity. In case of p-Pb collisions, the unfolding also includes a correction for fluctuations of the underlying background under the jet area. The order of these operations is not straightforward and it is studied in more details to estimate its systematic influence.

The flavour of the jet reconstructed in MC depends on the flavour of the quark, that fragmented to form this jet and it could be identified by different labeling methods. Such methods use the MC event history and they depend on the parton shower implementation. In this analysis, MC jets are considered as b-jets (coming from b quark fragmentation), if any B hadron is found in a cone $\Delta R<0.4$ around jet axis.
If no beauty hadron is present, but a charm hadron was found in this cone, the jet is labeled as c-jet. All other jets are considered as light-flavour jets, thus stemming from light quark or gluon fragmentations. 

The distribution of b-jets in simulations is used to estimate the tagging efficiencies.
The b-tagging efficiency 
is defined as the fraction of b-jets in the sample after applying selection cuts (tagged sample), divided by the number of b-jets without any selections. We also studied the purity of the MC tagged sample, given by the fraction of b-jets in the tagged sample.

The performance of the tagging algorithm is investigated via the mistagging rate, which is the efficiency of mistakenly tagging as beauty a jet originating from a c-quark or a light-flavour parton.
Mistagging rate studies are needed to extract a sample with high purity of b-jets. A high purity allows to suppress the contamination of light-flavour and charm jets, which is important since the measured fraction of b-jets to inclusive jets is at the level of 2-4\%, both in pp collisions at $\sqrt{s} = 7$ TeV \cite{CMS_frac} and in p-Pb collisions at  $\snn = 5.02$ TeV \cite{CMS_frac_pPb}, as measured by CMS experiment.

In this study, the transverse momentum of the jet at the particle level without detector effects, $\ptgen$, is used.

\label{jets}
\subsection{Track selection}
 In this performance study, charged tracks are reconstructed with simulated responses of the Inner Tracking System (ITS) and of the Time Projection Chamber (TPC) in a pseudorapidity region $|\eta|<0.9$ and transverse momentum $0.15<\pt<100$ GeV/$c$, are required to have at least 70 (out of 159) points in the TPC and at least one point in the Silicon Pixel Detector (SPD). Since some SPD regions are inefficient, the azimuthal distribution of these tracks is not completely uniform. This is why for the jet reconstruction additional tracks without reconstructed track points in the SPD are considered.
 
\subsection{Jet reconstruction and background estimation}

For the estimation of the background density and fluctuations in the jet reconstruction, a similar approach as it was applied by ALICE for charged jet measurements in p-Pb collisions \cite{ALICE_jetpPb} is used.
Namely, the anti-$k_{\mathrm{T}}$ algorithm from the FastJet package \cite{FastJet} with a resolution parameter of $R = 0.4$ is used to reconstruct charged jets. We use the CMS method \cite{CMS_bg}, which is suited for the more sparse environment of p-Pb collisions w.r.t Pb-Pb collisions, for the calculation of the background density
$$
\rho = \mathrm{median} \left( \frac{p_{\mathrm{T},i}}{A_i} \right)\cdot C,
$$
where $p_{\mathrm{T},i}$ is the transverse momentum of soft clusters, found by FastJet $k_{\mathrm{T}}$ algorithm, $A_i$ are the areas of these clusters and $C$ is the correction factor for the empty clusters.
The background fluctuations are estimated with a random cone (defined on the $\eta$-$\varphi$ plane) approach as
$$
\delta p_{\mathrm{T}}=\sum_i p_{\mathrm{T},i} - A_{cone}\cdot \rho,
$$
where the sum is over the track $p_{\mathrm{T}}$ in a cone and $A_{cone}$ is the cone area. For more details see \cite{ALICE_jetpPb}. Only jets with a jet axis in the pseudorapidity interval $|\eta_{\mathrm{jet}}|<0.5$ are considered, which ensures that the jet is fully located in the TPC acceptance.
In this study the background fluctuations are estimated with p-Pb events generated with HIJING with particle transport through the ALICE detector.
The background fluctuations under the jet area are found to be slightly smaller in the HIJING simulation than in the p-Pb data \cite{ALICE_jetpPb}.

\subsection{Correction for detector effects and background fluctuations}  
Corrections of jet momentum distribution for detector effects and background fluctuations are performed with MC simulations.
The detector response matrix is built by matching reconstructed jets at particle level without detector effects and jets at the detector level after particle transport through the ALICE detector.
The unfolding of the momentum distribution at detector level with inverse response matrix leads to the true jet momentum distribution. In this study, the Singular Value Decomposition (SVD) \cite{SVD} regularisation method is used.

After subtracting the constant background in each event in order to correct the jet transverse momentum $\ptjet$, one has to keep in mind that the background is not necessary constant, but may differ for different jets in a given event.
This is corrected statistically (not event by event) via an unfolding technique with a background fluctuation matrix $f(\delta \pt)$, where the background fluctuations $\delta \pt$ are described in Sec.~\ref{jets}. 
The actual unfolding is done with a matrix, which is the product of two matrices: the detector response matrix and the  background fluctuation matrix.

\begin{figure}[h]
\centering
\includegraphics[width=9cm,clip]{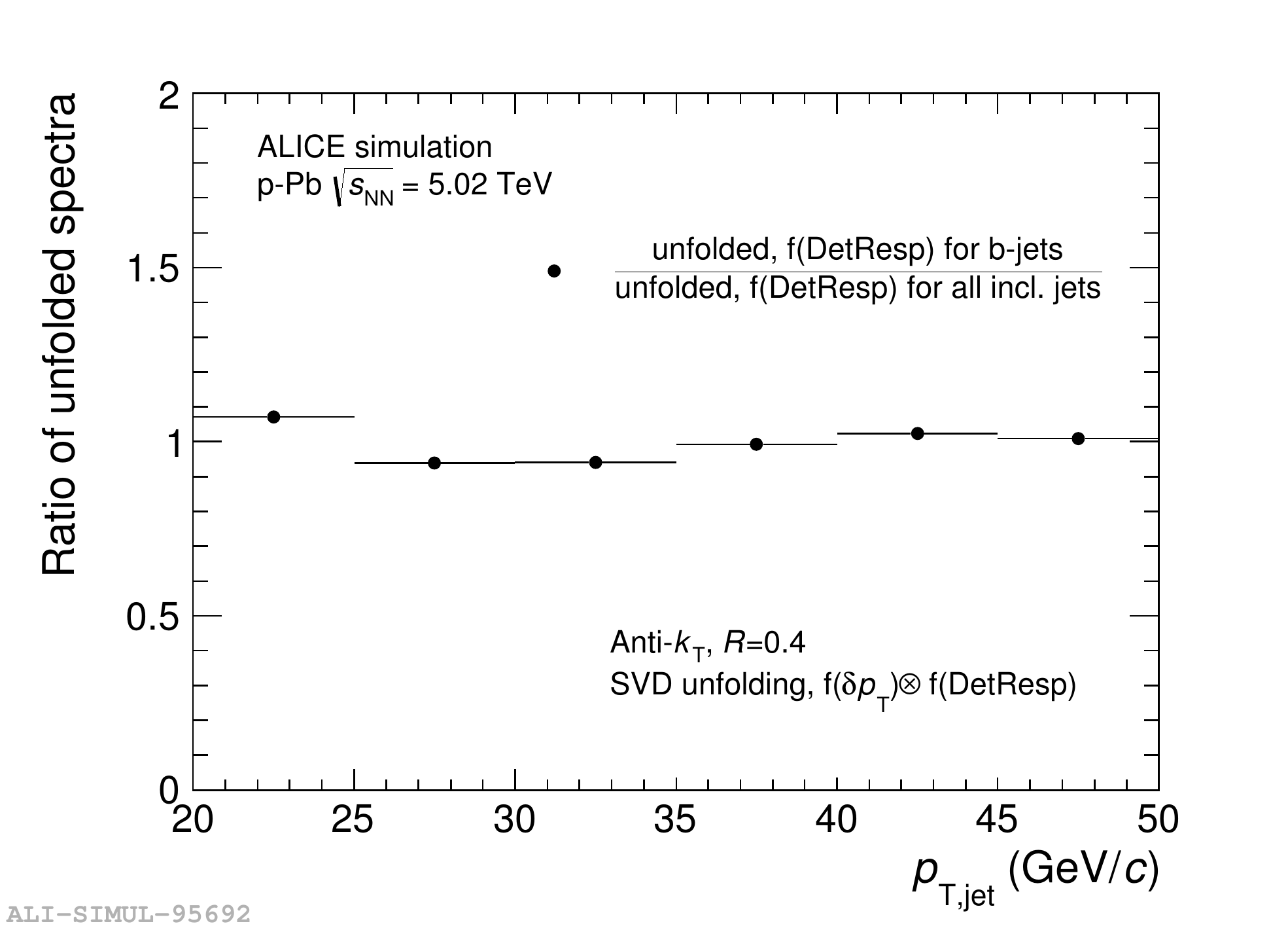}
\caption{Comparison of SVD unfolding of b-jet spectrum with two matrices: detector matrix for inclusive jets and for b-jets. Both matrices are combined with background fluctuation matrix from MC. The ratio of the two unfolded results is shown.}
\label{fig.Unf1}     
\end{figure}

The correction for detector effects is tested with two detector response matrices: the detector response for b-jets and the one for inclusive jets. Figure~\ref{fig.Unf1} shows the ratio of two unfolded b-jet $\pt$ spectra. In both cases the unfolding procedure yields similar results. This ensures that beauty and inclusive jets have similar detector response matrices and therefore the detectors response for inclusive jets can be used to unfold the tagged b-jet spectrum. 

\subsection{Secondary vertex tagging algorithm}
This section describes the performance of the secondary vertex b-tagging algorithm. This algorithm exploits the properties of beauty-hadron decays, characterised by displaced secondary vertices.  
First, all secondary vertices are reconstructed using tracks in the jet and  with transverse momentum $p_\mathrm{T, track}>1$ GeV/$c$. The secondary vertex is calculated as the point that minimizes the sum of the distances to the three tracks. For every jet, the secondary vertex, that is the furthermost from the primary vertex, is selected. Its properties are used to distinguish b-jets from light-flavour jets.

Discriminating variables in our b-jet tagging analysis are the secondary vertex dispersion and the significance of flight distance in a transverse plane.
The dispersion of the tracks in the vertex is defined as $\sigma_\mathrm{vtx}=\sqrt{d_1^2+d_2^2+d_3^2}$, where $d_{1,2,3}$ are the distances of the three tracks from the secondary vertex. This characterises the quality of the vertex reconstruction. The sign of the secondary vertex flight distance is defined w.r.t. the jet direction, so the signed length is $L_{xy}=\left| \vv{L'} \right| \mathrm{sign}(\vv{L'}\cdot \vv{p}_\mathrm{jet})$, where  $\vv{L'}$ is the vector between primary and secondary vertices. The significance of the flight distance in the transverse plane is then defined as $L_{xy}/\sigma_{L_{xy}}$, where $\sigma_{L_{xy}}$ is the uncertainty corresponding to $L_{xy}$. 
Cuts on these variables are applied on the selected (the furthermost) secondary vertex in the jet.

\begin{figure}[h]
\centering
\includegraphics[width=9cm,clip]{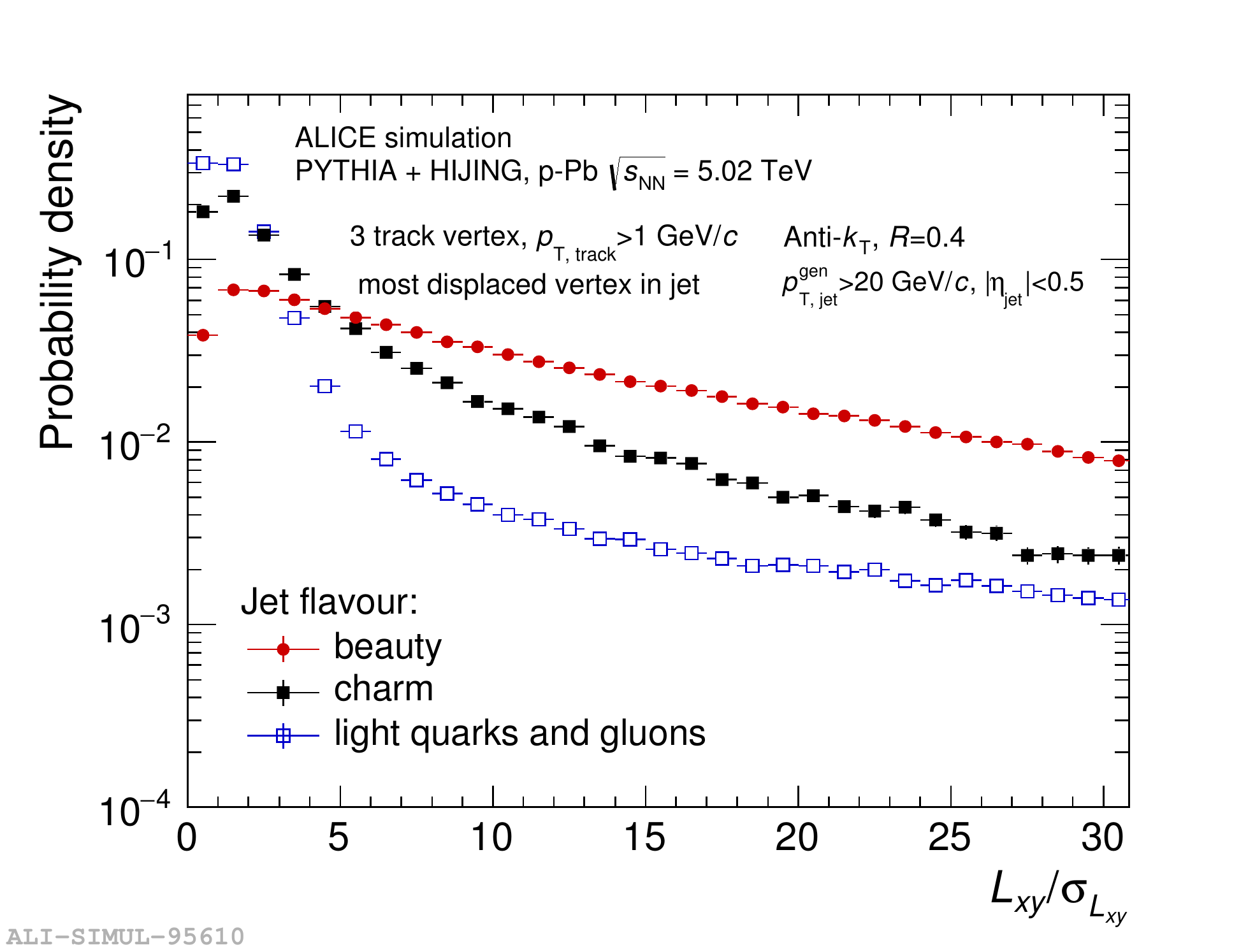}
\caption{Probability distribution of the signed flight distance significance of the most displaced secondary vertex, found in charged jets with $\ptgen > 20$ GeV/$c$ in p-Pb simulations at $\snn=5.02$ TeV. }
\label{fig.SLxy}      
\end{figure}

The distribution of the signed flight distance significance for jets with different flavours with transverse momentum \mbox{$\ptgen>20$ GeV/$c$} is shown in Fig.~\ref{fig.SLxy}. 
Its discriminating power is manifested by differences in shapes of its distribution for b-jets and for light and charm jets. Consequently, the larger the cut value of $L_{xy}/\sigma_{L_{xy}}$, the more light and charm jets are rejected compared to beauty jets. 

\section{Performance of secondary vertex b-tagging algorithm}
Different values of cuts on discriminating variables ($L_{xy}/\sigma_{L_{xy}}$ and $\sigma_\mathrm{vtx}$) yield different algorithm performance. Our goal is to find the working point with sufficiently hight b-jet purity  with the mistagging rate of light jets about 100 times smaller than tagging efficiency to reduce background, as discussed in Sec.~\ref{sec1}.

\begin{figure}[h]
\centering
\includegraphics[width=7.5cm,clip]{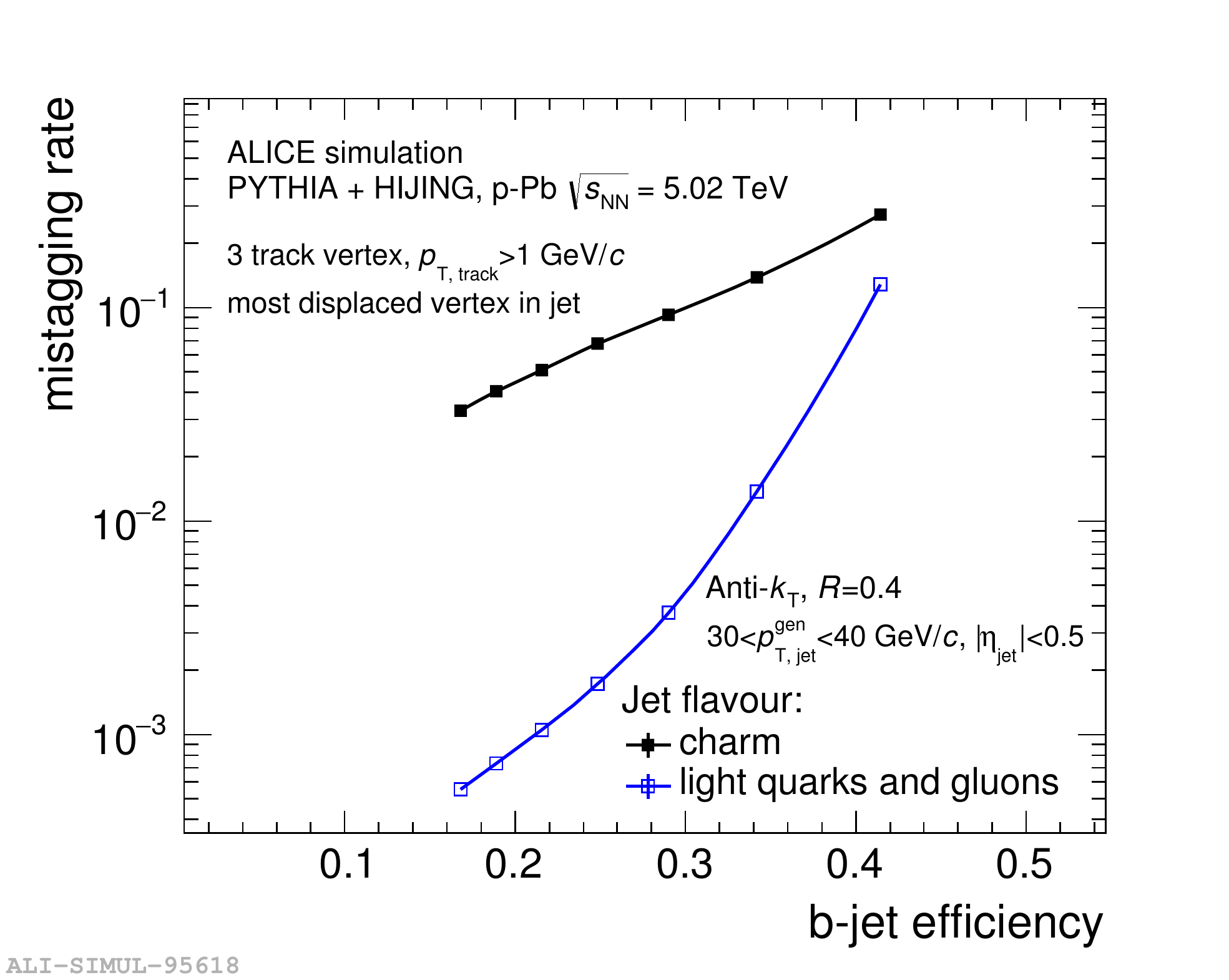}
\includegraphics[width=7.5cm, clip]{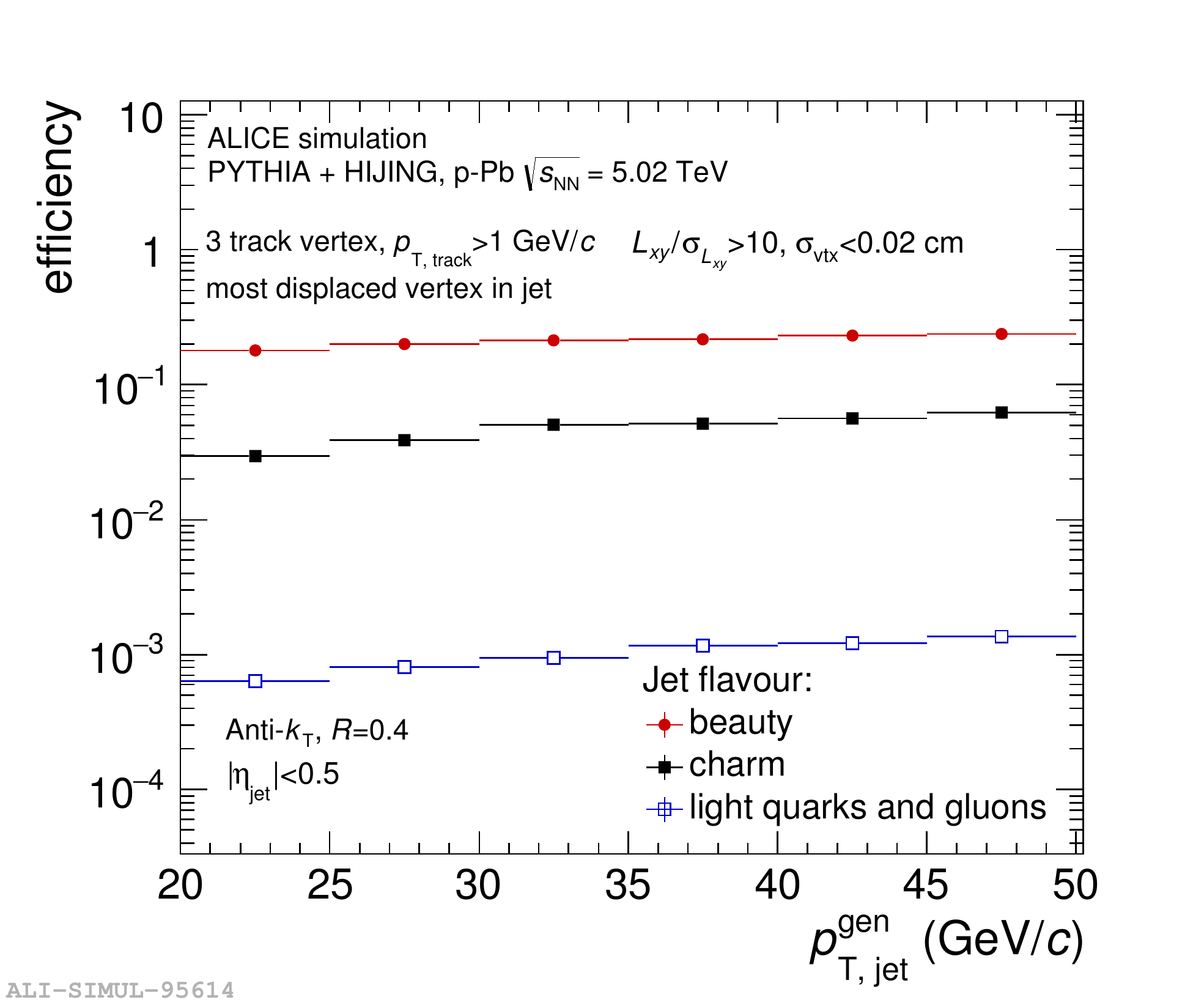}
\caption{Left: Mistagging rate vs the b-tagging efficiency of SV tagging algorithm for different operating points for jet transverse momentum range $30<\ptgen<40$ GeV/$c$. Right: The $\ptgen$ dependence of the b-tagging efficiency and mistagging rate for $L_{xy}/\sigma_{L_{xy}}>10$ and $\sigma_\mathrm{vtx}<0.02$ cm.}
\label{fig.Eff}       
\end{figure}

Figure~\ref{fig.Eff}, left, shows the tagging performance (mistagging rate vs the b-tagging efficiency) in jet $\ptgen$ range $30<\ptgen<40$ GeV/$c$ for different operating points.
These are obtained by varying the cuts on $L_{xy}/\sigma_{L_{xy}}$ (from 2 to 14), while the cut on $\sigma_\mathrm{vtx}<0.02$ cm is fixed.
In the considered region of the jet $\ptgen$, efficiencies are almost constant.
Looser cuts result in a larger tagged sample and higher tagging efficiency, but also higher mistagging rate, and therefore reduce the purity of the sample.

The tagging and mistagging efficiencies
at particle level for the chosen operating point with $L_{xy}/\sigma_{L_{xy}}>10$ and $\sigma_\mathrm{vtx}<0.02$ cm are reported in Fig.~\ref{fig.Eff}, right panel, as a function of the jet transverse momentum $\ptgen$. The b-tagging efficiency is around 20$\%$, while the efficiency to tag light-flavour jets is about two orders of magnitude lower and the efficiency to tag charm jets is about 3 to 5 times lower than the b-tagging efficiency.

\begin{figure}[h]
\centering
\includegraphics[width=9cm,clip]{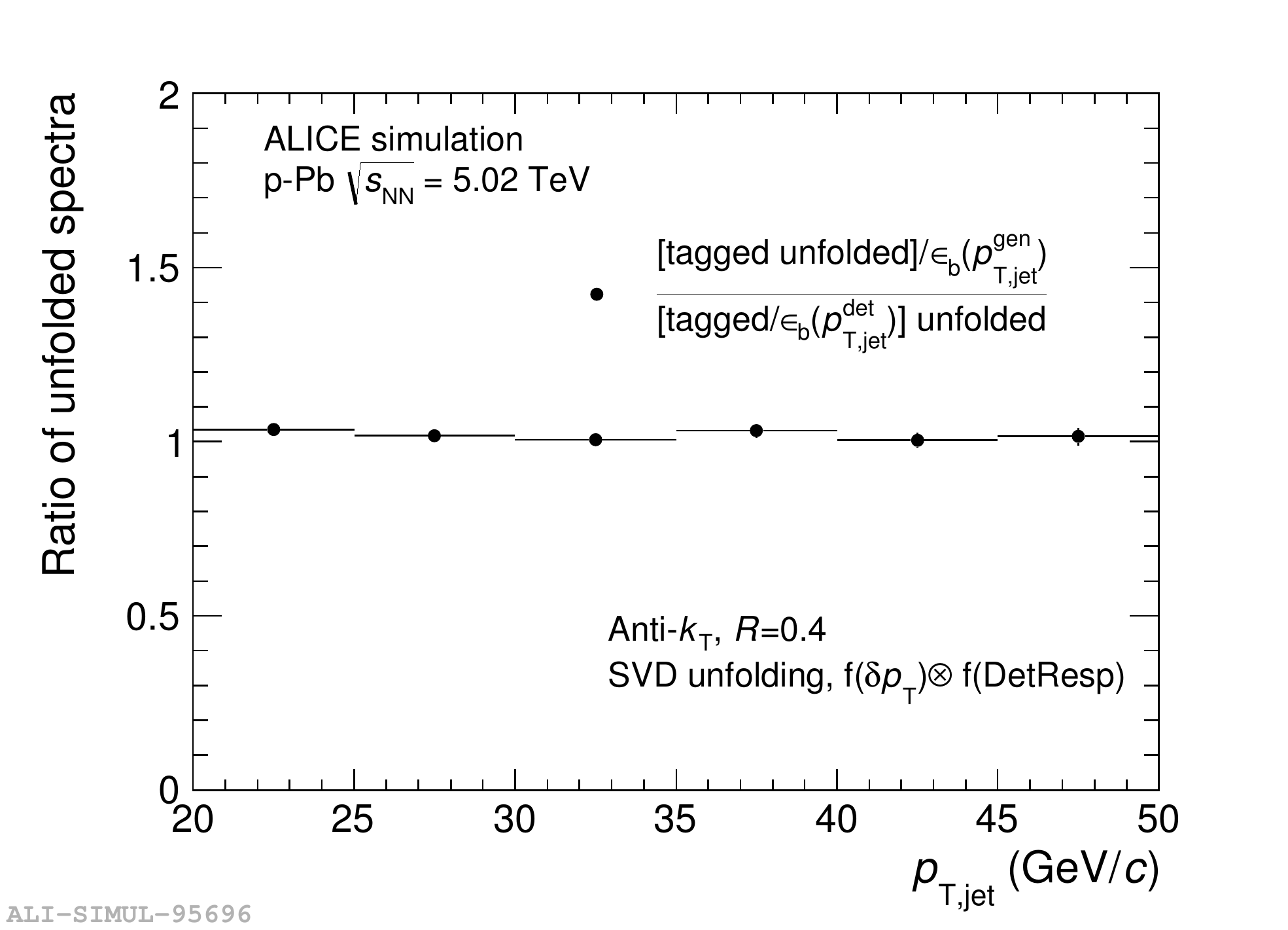}
\caption{Comparison of two sequences of corrections: SVD unfolding and correction for tagging performance and vice versa. The ratio of the two corrected spectra is shown.}
\label{fig.Unf2}       
\end{figure}

The systematic influence of the order of unfolding and corrections on purity and efficiency is studied. The stability is tested for two scenarios:
\begin{itemize}
\item the tagged b-jet spectrum is first unfolded for detector response and background fluctuations, then corrected for the b-tagging efficiency as a function of $\ptgen$ (at particle level);
\item the tagged b-jet spectrum is first corrected for the b-tagging efficiency as a function of $p_\mathrm{T}^\mathrm{det}$ and then unfolded for detector response and background fluctuations.  
\end{itemize}  
Figure~\ref{fig.Unf2} displays comparison of the spectra obtained by the two scenarios. As it can be seen, they are in good agreement with each other, suggesting that both scenarios can be considered in p-Pb collisions.

\section{Summary}
The study of the performance of the b-jet tagging algorithm based on displaced secondary vertices  with MC simulation of p-Pb events for ALICE detector in the jet transverse momentum range $20<\ptgen<50$ GeV/$c$ was presented.

As discriminating variables, the secondary vertex dispersion and the significance of flight distance in the transverse plane were chosen.
The studied selections, based on the discriminating variables, translate into a b-jet efficiency of 20\%, suppressing light-flavour jets contamination by two orders of magnitude. The tagging purity itself is not discussed here. 

Corrections for the detector response and background fluctuations were studied.
It was found that the b-jet spectrum can be corrected with a detector response matrix for inclusive jets.
Furthermore, the order in which the unfolding and the purity and the efficiency corrections are applied was found to have a negligible influence on systematics.

Other ways to reduce the background are being studied, for example the rejection  of tracks with V$^0$ topology. 
To stabilize the performance of the algorithm, the study of different selections for different jet transverse momentum is ongoing, as well as the estimation of the purity with data-driven methods and with different MC generators.


\section{Acknowledgment}
This work was supported by grant number LG15052 of Ministry of Education, Youth and Sports of Czech Republic and by Grant Agency of the Czech Technical University in Prague, grant No. SGS16/242/OHK4/3T/14.


\end{document}